\documentclass[amsmath,amssymb,superscriptaddress,aps,a4paper,prb,twocolumn]{revtex4} 

\newcommand{\iga}[2]{In\ensuremath{_{#1}}Ga\ensuremath{_{#2}}As}
\newcommand{\iaa}[2]{In\ensuremath{_{#1}}Al\ensuremath{_{#2}}As}

\renewcommand{\eqref}[1]{Eq.\ \ref{#1}}

\usepackage[dvips]{color}
\usepackage{graphicx}
\usepackage{amsmath}
\usepackage{amssymb}
\usepackage{amsfonts}
\usepackage{amsthm}
\usepackage{textcomp}
\usepackage{natbib}
\usepackage{gensymb} 
\usepackage[seperr, load={}]{siunitx}

\begin{document}

\title{\textbf{Proximity effect in a two dimensional electron gas probed with a lateral quantum dot}}

\author{F. Deon}
	\email{f.deon@sns.it}
	\affiliation{NEST, Istituto Nanoscienze-CNR  and Scuola Normale Superiore, I-56127 Pisa, Italy}
\author{V. Pellegrini}
	\affiliation{NEST, Istituto Nanoscienze-CNR  and Scuola Normale Superiore, I-56127 Pisa, Italy}\date{\today}
\author{F. Giazotto}
	\affiliation{NEST, Istituto Nanoscienze-CNR  and Scuola Normale Superiore, I-56127 Pisa, Italy}\date{\today}
\author{G. Biasiol} 
		\affiliation{CNR-IOM, Laboratorio TASC, Area Science Park, I-34149 Trieste, Italy}
\author{L. Sorba}
	\affiliation{NEST, Istituto Nanoscienze-CNR  and Scuola Normale Superiore, I-56127 Pisa, Italy}
\author{F. Beltram}
	\affiliation{NEST, Istituto Nanoscienze-CNR  and Scuola Normale Superiore, I-56127 Pisa, Italy}

\date{\today}      

\begin{abstract}
We report low-temperature transport measurements performed on a planar Nb-InGaAs-Nb proximity Josephson junction hosting a gate-defined lateral quantum dot in the weak-link. We first study quasiparticle and Josephson transport through the open junction, when all gates are grounded. When the quantum dot is defined in the normal region by electrostatic depletion, cotunneling spectroscopy allows us to directly probe the energy gap induced in the two dimensional electron gas. Our data show good qualitative agreement with a model describing resonant tunneling through an Anderson impurity connected to superconducting electrodes. These results demonstrate the feasibility of top-gated nanodevices based on a two-dimensional electron gas coupled to a superconductor.
\end{abstract}

\maketitle

Mesoscopic devices obtained by coupling the two-dimensional electron gas (2DEG) confined in semiconductor heterostructures to a superconductor have attracted considerable interest, both in the context of device applications\cite{Akazaki1996}, and from the point of view of basic research\cite{Golubov2004}. In fact, the high mobilities that can be achieved in modulatio-doped heterostructures make these materials suitable for experimentally studying the proximity effect\cite{deGennes1999} in ballistic two-dimensional normal conductors. Moreover, the possibility of patterning the 2DEG or introducing lateral confinement in top-gated devices opens the way to a class of devices in which the proximity effect coexists with quantum confinement and reduced dimensionality\cite{Bauch2005,Deon2011}, or is affected by phase coherence\cite{Dolcini2007}.
More recently, superconductor-coupled 2DEGs have been proposed\cite{Alicea2010} as a promising material system for the experimental realization of a topological superconductor, and to explore the existence of Majorana bound states in the solid state\cite{Wilczek2009}.
Most of the work performed to date on superconductor-coupled 2DEGs exploited heterostructures based on InAs\cite{Giazotto2004} or \iga{x}{1-x} alloys\cite{Schapers1997} with molar fraction $x\geq 0.75$, usually in association with superconducting Nb: these low-bandgap semiconductors are characterized by a small effective mass, large spin-orbit coupling constant and gyromagnetic factor, and form negligible Schottky barriers with most metals. While this last property is crucial for the preparation of highly transparent superconductor-2DEG junctions, the difficulty of obtaing stable gated devices with these materials has somehow hampered the study of gated 2DEG hybrid systems. 

We have recently reported\cite{Deon2011} the realization of a hybrid device which combines a planar Nb-InGaAs-Nb proximity Josephson junction with a gate-defined lateral quantum dot (QD), obtained by means of top-down nanofabrication starting from an \iga{0.8}{0.2}/\iaa{0.75}{0.25} 2DEG. 
The reported procedure offers a top-down nanofabrication alternative to the realization of hybrid QD devices (see [\onlinecite{DeFranceschi2010}] and refs. therein), based sofar on self-assembled nanostructures. 
We showed that the QD can be tuned to the regimes of weak and strong coupling, and can be used as a direct probe of the proximity effect in the 2DEG. In the present work we report a detailed analysis of quasiparticle and Josephson transport in the open junction configuration (all gates set to ground), and study the evolution of the elastic cotunneling conductance through the QD as a function of magnetic field and temperature. Measurements are compared with a theoretical model describing resonant tunneling across an Anderson impurity coupled to superconducting electrodes, with good qualitative agreement.

\begin{figure} \label{fig1}
\includegraphics{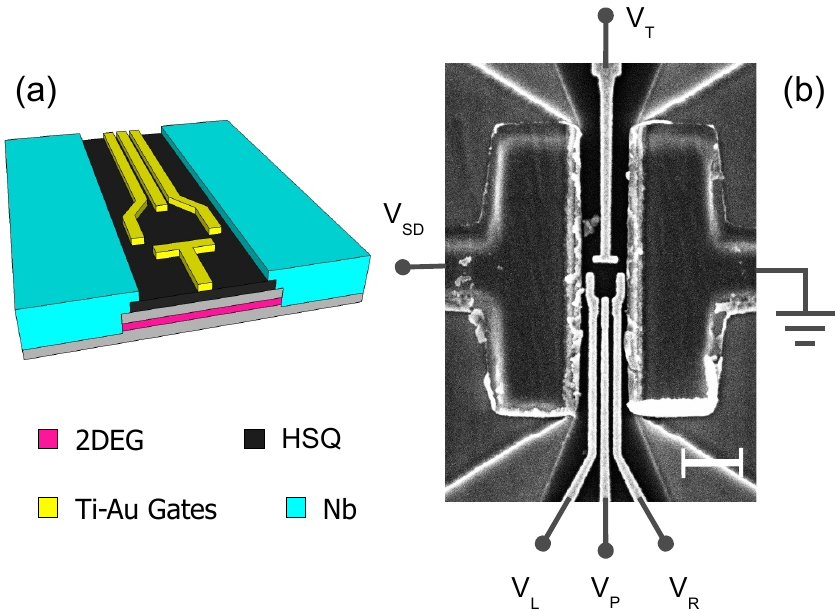}
\caption{(Color online) (a) Graphical rendering of Nature Nanotech.the device layout: the quantum dot is defined by electrostatic gates placed within the normal region of a planar Nb-InGaAs 2DEG-Nb junction. A \SI{60}{nm}-thick hydrogen silsesquioxane (HSQ) layer provides gate insulation from the 2DEG. (b) Scanning electron micrograph of a representative device. Scale bar is \SI{650}{nm}.}
\end{figure}

A cartoon and a scanning electron micrograph of a representative device are shown in Fig. 1. We start from a metamorphic \iga{0.8}{0.2}/\iaa{0.75}{0.25} epitaxial heterostructure, grown on a (001) GaAs substrate. The 2DEG is confined in a \SI{15}{nm}-thick quantum well placed \SI{45}{nm} from the surface. Electron density and mobility are $n_s=5.9\cdot 10^{11}\SI{}{cm^{-2}}$ and $\mu=1.8\cdot 10^{5}\SI{}{cm^2/Vs}$, respectively. The electron effective mass is $\sim 0.03m_0$, where $m_0$ is the free electron mass\cite{Desrat2004}. A planar proximity Josephson junction is obtained by laterally contacting a strip defined by electron beam lithography and wet etching in the 2DEG. Junction length and width are $L=\SI{650}{nm}$, and $W=\SI{3}{\micro m}$, respectively. Furthermore, we insert a \SI{60}{nm}-thick HSQ film to provide insulation of the electrostatic gates from the underlying semiconductor. Further details regarding the heterostructure and sample fabrication are reported elsewhere\cite{Deon2011}.

\begin{figure} \label{fig2}
\includegraphics{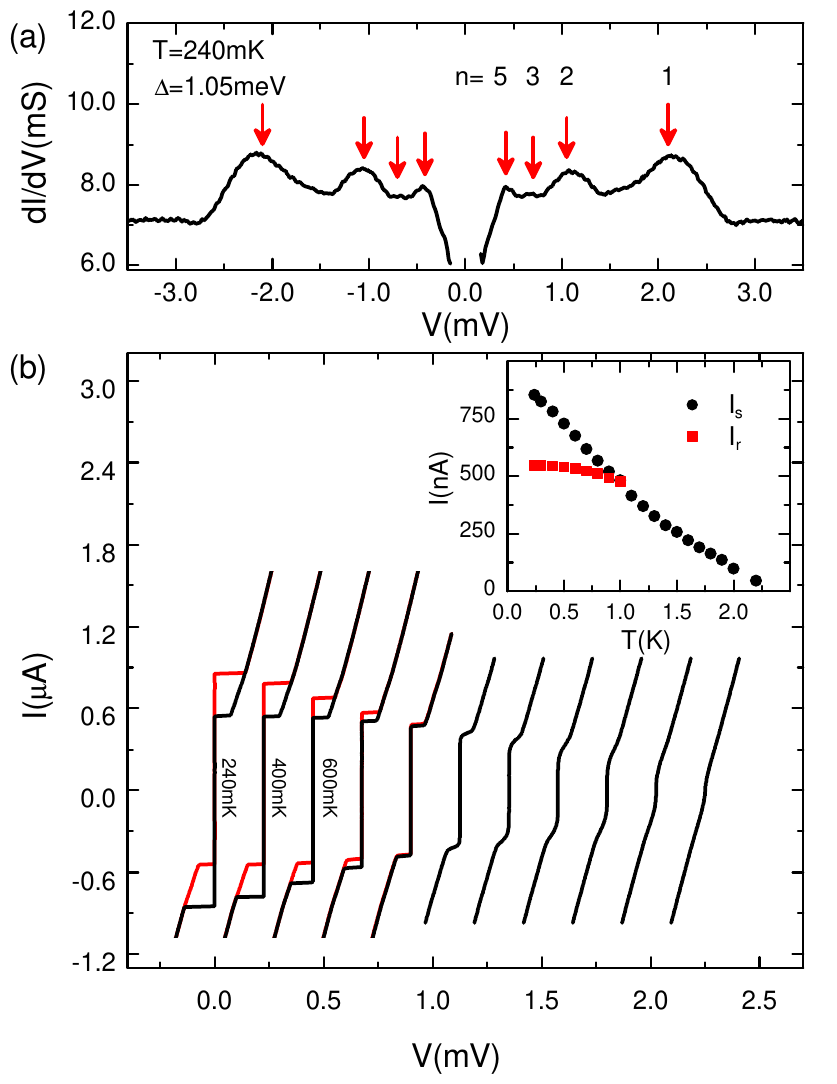}
\caption{(Color online) (a) Differential conductance versus source-drain voltage $V$ at $T=\SI{240}{mK}$. Red arrows are placed at voltages $V=2\Delta/en$, with $\Delta=\SI{1.05}{meV}$, and coincide with $dI/dV$ peaks linked to multiple Andreev reflections (MARs).
(b) Low-bias voltage-current characteristics of the junction at \SI{240}{mK}, and between \SI{400}{mK} and \SI{2}{K}, with temperature intervals of \SI{200}{mK}. The curves are horizontally offset for clarity. The temperature dependence of the switching and retrapping currents ($I_s$ and $I_r$, respectively) is displayed in the inset.}
\end{figure}

We first address the properties of the Nb-2DEG-Nb proximity junction in the open configuration. Measurements in this configuration are performed in a filtered $^3$He cryostat, down to \SI{240}{mK}.
The differential conductance at source-drain voltage bias up to $|V|\sim3\Delta/e$ is plotted in Fig. 2(a). At least three orders of multiple Andreev reflection (MAR)\cite{Flensberg1988} are clearly visible: the $dI/dV$ peak positions approximately match the values $V_n=2\Delta/en$, with $n=1,2,3,5$ and $\Delta=\SI{1.05}{meV}$, marked by the red arrows in Fig. 2(a). The Thouless energy, which in the clean limit reads $\epsilon_{Th}=\hbar v_F/L$, is $\sim \SI{750}{\micro eV}$, so that our junction falls in the intermediate-length regime. Finally, the junction length $L$ is smaller than the mean free path $l_p=\SI{2.3}{\micro m}$, so that electron transport in the 2DEG strip is ballistic.

In Fig. 2(b) we show voltage-current characteristics of the junction between \SI{240}{mk} and \SI{2.2}{K}. In the low temperature range $T\lesssim\SI{1}{K}$ the curves are hysteretic: as $I$ is raised above a critical value $I_s$, the junction switches to the dissipative branch of the I-V characteristic, whereas retrapping into the dissipationless branch occurs at a lower value $I_r$ of $I$\cite{Courtois2008}.
For $T\gtrsim\SI{1}{K}$ the hysteresis disappears: in this temperature range we take as the experimental switching current the value of $I$ for which $dV/dI=R_n/2$, with $R_n$ defined\cite{Dubos2001} as the differential resistance at $I=\SI{900}{nA}$\cite{note1}.
The values of $I_s$ and $I_r$ are displayed in the inset to Fig. 2. Since the switching current does not saturate down to \SI{250}{mK}, we estimate\cite{Hammer2007} an upper bound $\Delta^*<\SI{20}{\micro eV}$ for the minigap of the junction in the open configuration. This value is considerably smaller than $\epsilon_{Th}$, a fact that can be attributed to the nonideal transmissivity of the Nb-2DEG interfaces and to the geometry of our weak-link: in fact, the presence of large portions of 2DEG outside the junction region can introduce subgap states in the 2DEG, by inverse proximity effect\cite{Giazotto2004,Heikkila2002}.

\begin{figure} \label{Fig3}
\includegraphics{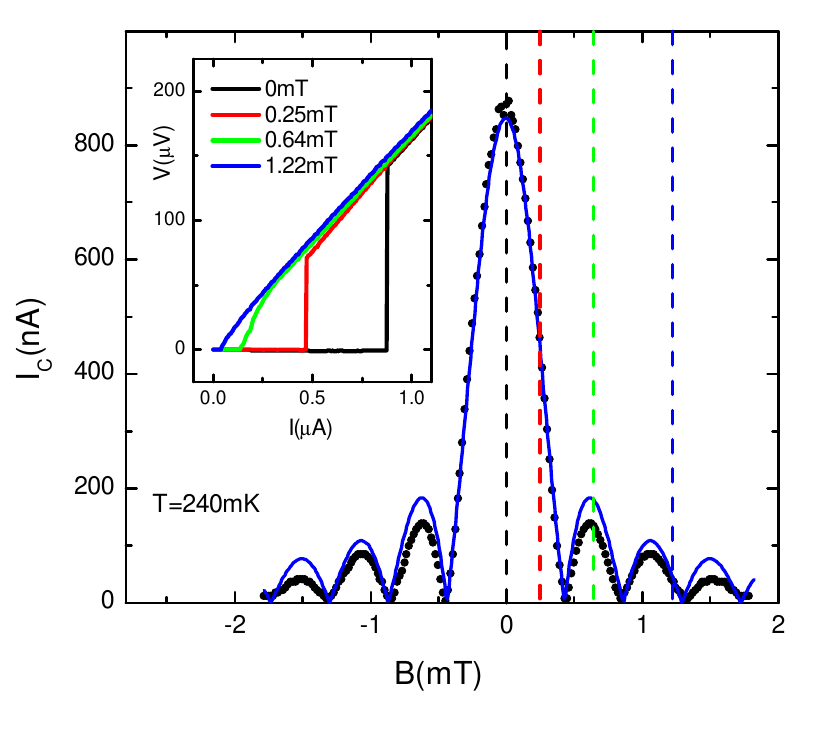}
\caption{(Color online) Plot of the switching current $I_s$ as a function of orthogonal magnetic field $B$ (black dots) in the open juction configuration. The blue line corresponds to the best fit to the data of the Fraunhofer-like interference pattern (see text). Voltage-current characteristic at representative values of $B$ (marked by dashed lines in the plot) are shown in the inset.}
\end{figure}

In Fig. 3(b) we show the magnetic interference pattern (MIP) of the open junction, i.e. the evolution of $I_s$ as a function of magnetic field $B$ applied normal to the junction plane (experimental points are shown as black dots). 
Representative traces along the MIP are shown in the inset. As in the case of conventional Josephson tunnel junctions in parallel field, the MIP is a manifestation of phase modulation in the interior of the superconducting leads, induced by the applied magnetic field\cite{Tinkham1996}. 
The continuous blue line is the best fit to the data of the expression $I_s=I_s^0\left|\mathrm{sin}(\pi\alpha LWB/\Phi_0)/(\pi\alpha LWB/\Phi_0)\right|.$ Here $\Phi_0=h/2e$ is the magnetic flux quantum, $I_s^0$ the zero-field switching current, and $\alpha$ a dimensionless parameter required to reproduce the observed periodicity in the applied field. The latter accounts both for the effect of field focussing due to expulsion from the Nb leads\cite{Gu1979} (Meissner effect), and for possible deviations from the usual $\Phi_0$ periodicity, which are expected for planar ballistic junctions with a finite $L/W$ ratio\cite{Heida1998,Barzykin1999}.
From the fit procedure we extract $\alpha\sim2.5$. However, the measurement of the MIP on a single junction does not allow to separately quantify the contributions of these two effects. The observed closeness of the experimental points to the Fraunhofer-like pattern might be unexpected, for several reasons: (i) the current-phase relation of a ballistic proximity junction at low temperatures is expected to be sensibly anharmonic\cite{Golubov2004, Grajcar2002}; (ii) the phase periodicity of the MIP nodes and the MIP functional profile are expected to deviate from the Fraunhofer formula in the case of junctions with a finite $L/W$ ratio\cite{Heida1998}; (iii) field expulsion from the Nb contacts could lead in our geometry to nonuniform flux density through the normal region.
Finally, the almost complete suppression of $I_s$ at the nodes (Fig.3) suggests a high degree of uniformity of the critical current density along our junction.

\begin{figure} \label{fig4}
\includegraphics{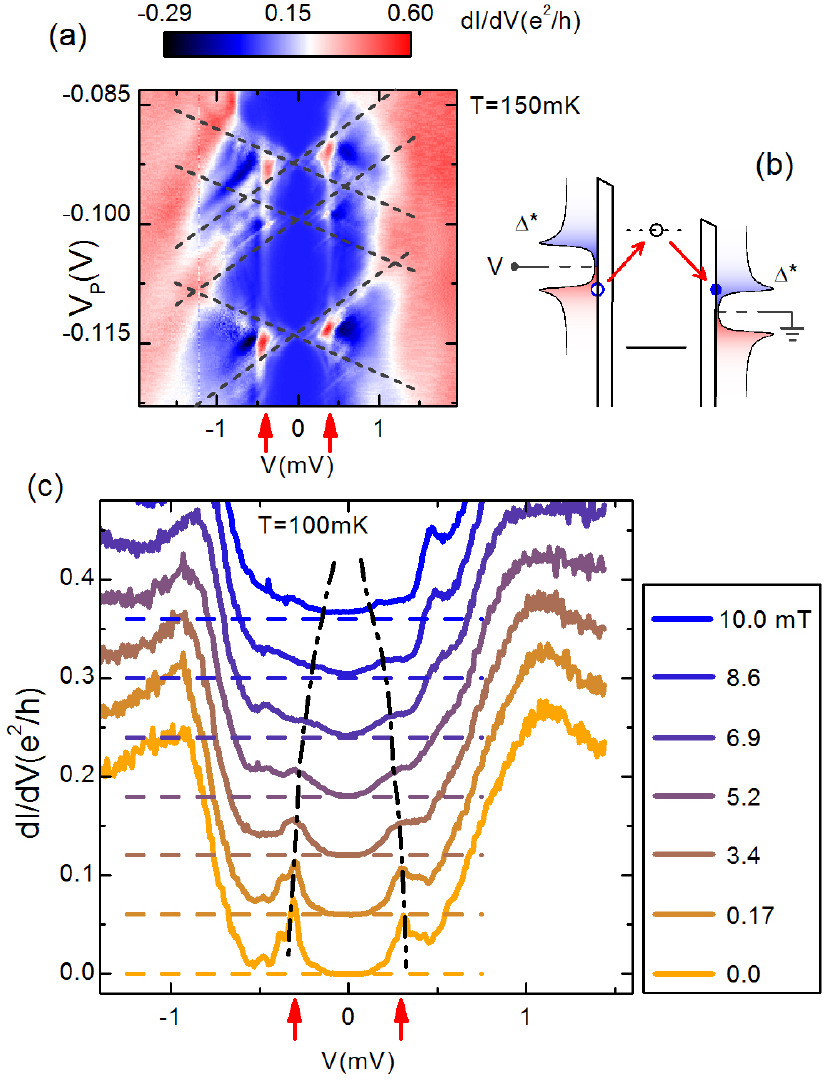}
\caption{(Color online) (a) Stability plot for the hybrid QD in the regime of strong coupling to the leads. The arrows point to elastic cotunneling peaks, at constant energy $2\Delta^*$. The energy diagram in (b) illustrates one of such cotunneling processes. (b) Magnetic field dependence of the $dI/dV$ curves for $V_P$ corresponding to the middle of a Coulomb valley: proximity-induced superconductivity is suppressed in a perpendicular critical field $B\simeq\SI{10}{mT}$. The curves are vertically offset for clarity (dashed lines indicate the position of $dI/dV=0$ for each curve). Dot-dashed lines are a guide to the eye, marking the evolution of peak positions.}
\end{figure}

When the top gates are polarized, confining a QD in the weak-link, the system can be viewed as a S-2DEG-QD-2DEG-S junction. The S-2DEG junctions can be considered to be at equilibrium, with the applied voltage bias dropping entirely across the QD. By tuning the values of $V_L$, $V_R$ and $V_T$ (see Fig. 1b) it is possible to vary the transparency of the QD-2DEG barriers, while the voltage $V_P$ applied to the fourth gate is used to tune the charge state of the QD. The distance between the QD barriers and the S-2DEG interfaces is of the order of $\sim$\SI{100}{nm}, which yields a Thouless energy for each of the S-2DEG junctions of $\sim\SI{4.9}{meV}$\cite{note2}. In the short junction limit $\Delta$ sets an upper bound to the the proximity-induced gap $\Delta^*$, while the actual value could be considerably suppressed by the presence of normal scattering at the S-2DEG interface.

Measurements discussed in the following are performed in a filtered $^3$He/$^4$He dilution refrigerator.
In Fig. 4(a) we show the stability plot (i.e., the differential conductance versus $V$ and $V_P$) for the hybrid QD in the regime of strong coupling to the leads, measured at \SI{150}{mK}. Higher order cotunneling processes contribute to electrical transport through the QD inside Coulomb diamonds, where single-electron tunneling is suppressed by Coulomb blockade. Symmetric peaks in $dI/dV$, located at constant source drain voltages $\pm\SI{330}{\micro eV}$ irrespective of the QD charge state, are indicated by the red arrows in Fig 4(a,c). They are related to the onset at $V=2\Delta^*/e$ of elastic cotunneling processes such as the one depicted in the energy diagram of Fig 4 (b) (with $\Delta^*\simeq\SI{165}{\micro eV}$). The proximity effect in the 2DEG is suppressed when a small magnetic field is applied perpendicular to the junction plane. In Fig. 4(c) we plot the evolution of the $dI/dV$ traces at a fixed value of $V_P$, corresponding to the middle of a Coulomb valley. As B increases from $0$ to \SI{10}{mT}, the intensity of the elastic cotunneling peaks decreases, and their position collapses to smaller energies (with the appearance of nonzero differential conductance at zero bias). This can be regarded as a measurement of the critical field of the proximized 2DEG regions.

\begin{figure} \label{fig5}
\includegraphics{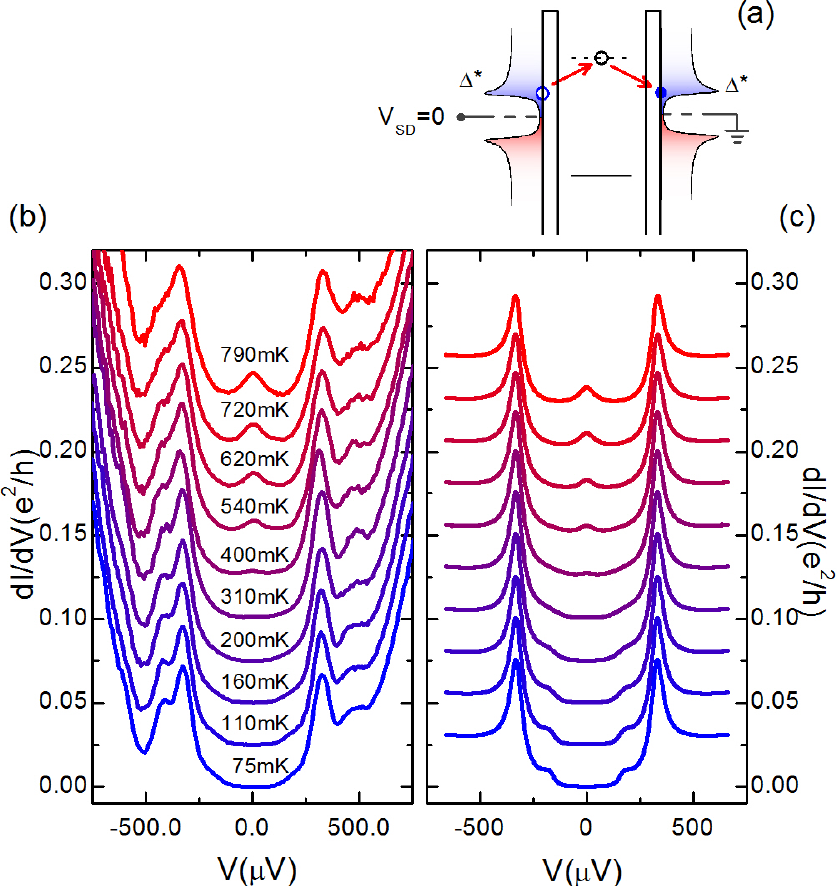}
\caption{(Color online) Temperature dependence of the $dI/dV$ vs $V$ curves at a fixed value of $V_P$ corresponding to the middle of a Coulomb valley: at higher temperatures a peak develops at $dI/dV=0$, where the alignment of the DOS peaks enhances the contribution of second-order elastic processes involving thermally-activated quasiparticles [see diagram in (a)]. Qualitatively similar behavior is observed in all diamonds. (b) Experimental data. (c) Differential conductance curves calculated using Eq. (1), for the same $T$ values as in panel (b). The DOS in the two leads was assumed to follow a broadened BCS-like profile (see text). In both panels the curves are vertically offset by $0.025\,e^2/h$ for clarity.}
\end{figure}

In Fig. 5(b) we show the experimental $dI/dV$ at zero magnetic field, bath temperatures between \SI{75}{mK} and \SI{790}{mK}, and a fixed value of $V_P$ corresponding to the middle of a Coulomb valley (the same qualitative behavior is observed in all diamonds). When the temperature is increased, a conductance peak with growing intensity developes around zero bias. In fact, as soon as $T$ becomes comparable with $\Delta^*/k_B$, quasiparticles are thermally excited across the proximity-induced gap, and elastic cotunneling processes like the one depicted in Fig. 5(a) contribute to transport at small bias voltages. The contribution of such processes to the differential conductance is largest at $V=0$, where the peaks in the 2DEG density of states are aligned. A similar behavior is known to take place in conventional superconductor-insulator-superconductor junctions\cite{Tinkham1996}, although in that case second-order cotunneling processes are replaced by direct tunneling across the insulating layer. Note the independence in our experiment of $\Delta^*$ on the bath temperature up to $T=\SI{790}{mK}$ (where $k_BT\sim\SI{68}{\micro eV}$): this is consistent with the high value of the Thouless energy and with the negligible variation of $\Delta$ in the explored temperature interval, which is much lower than the critical temperature of the Nb films.
In Fig 5(c) we plot conductance curves calculated using a model developed in Ref. [\onlinecite{Yeyati1997}], which describes a QD coupled to BCS superconductors. The QD is modeled as an Anderson impurity with resonant level energy $\epsilon$. In the limit of strong Coulomb interaction, MAR processes across the QD can be neglected, and the current is given by

\begin{eqnarray*}
 I(V) & = & \frac{4e}{h}\int_{-\infty}^{+\infty} \mathrm{d}\omega\frac{\Gamma^S_L(\omega)\Gamma^S_R(\omega)}{(\omega-\epsilon)^2+[\Gamma^S_L(\omega)+\Gamma^S_R(\omega)]^2}\times \\
 & \times & \left[n_F(\omega-eV/2)-n_F(\omega+eV/2)\right]. \qquad\qquad (1) \\
\end{eqnarray*}

Here, $n_F(\omega)= (1+e^{\omega/k_BT})^{-1}$ is the Fermi-Dirac distribution function and $\Gamma_{L(R)}^S(\omega)=\Gamma_{L(R)}\cdot\rho^S_{L(R)}(\omega)$, where $\rho^S_{L(R)}(\omega)$ is the electron DOS in the left (right) contact and $\Gamma_{L(R)}$ the electron tunneling rate for the left (right) barrier. The curves in Fig. 5(b) are calculated assuming that the local DOS in the proximized 2DEG can be described with a smeared BCS-like DOS\cite{Dynes1978} $\rho^S_L(\omega)=\rho^S_R(\omega)=|\Re\mathrm{e}[(\omega+i\gamma)/[(\omega+i\gamma)^2-{\Delta^*}^2]^{1/2}]|$,
with $\Delta^*=\SI{165}{\micro eV}$ and broadening parameter $\gamma/\Delta^*=10^{-1}$. Moreover, we assumed $\epsilon=7\Delta^*$ and $\Gamma_L=\Gamma_R=0.57\Delta^*$. 
Although such a simple lineshape does not allow for the accurate description of the DOS in the 2DEG, the main qualitative features of the data in Fig. 5(b) and their temperature dependence are well captured by the model.

In summary, we have reported the experimental investigation of electron transport in a hybrid device in which top gates allow to define a lateral QD in the normal region of a planar Nb-2DEG-Nb junction. We explored quasiparticle and Josephson transport through the open junction, and reported the evolution of the elastic cotunneling conductance through the QD as a function of magnetic field and temperature. Good qualitative agreement was found with a model describing resonant tunneling through an Anderson impurity.

The authors acknowledge partial financial support from the E.U. Projects HYSWITCH (grant No. FP6-517567) and MICROKELVIN (grant No. FP7-228464), the MIUR-FIRB No. RBIN06JB4C.


\begin{thebibliography}{24}
\expandafter\ifx\csname natexlab\endcsname\relax\def\natexlab#1{#1}\fi
\expandafter\ifx\csname bibnamefont\endcsname\relax
  \def\bibnamefont#1{#1}\fi
\expandafter\ifx\csname bibfnamefont\endcsname\relax
  \def\bibfnamefont#1{#1}\fi
\expandafter\ifx\csname citenamefont\endcsname\relax
  \def\citenamefont#1{#1}\fi
\expandafter\ifx\csname url\endcsname\relax
  \def\url#1{\texttt{#1}}\fi
\expandafter\ifx\csname urlprefix\endcsname\relax\def\urlprefix{URL }\fi
\providecommand{\bibinfo}[2]{#2}
\providecommand{\eprint}[2][]{\url{#2}}

\bibitem[{\citenamefont{Akazaki et~al.}(1996)\citenamefont{Akazaki, Takayanagi,
  Nitta, and Enoki}}]{Akazaki1996}
\bibinfo{author}{\bibfnamefont{T.}~\bibnamefont{Akazaki}},
  \bibinfo{author}{\bibfnamefont{H.}~\bibnamefont{Takayanagi}},
  \bibinfo{author}{\bibfnamefont{J.}~\bibnamefont{Nitta}}, \bibnamefont{and}
  \bibinfo{author}{\bibfnamefont{T.}~\bibnamefont{Enoki}},
  \bibinfo{journal}{Appl. Phys. Lett.} \textbf{\bibinfo{volume}{68}},
  \bibinfo{pages}{418} (\bibinfo{year}{1996}).

\bibitem[{\citenamefont{Golubov et~al.}(2004)\citenamefont{Golubov, Kupriyanov,
  and Il'ichev}}]{Golubov2004}
\bibinfo{author}{\bibfnamefont{A.~A.} \bibnamefont{Golubov}},
  \bibinfo{author}{\bibfnamefont{M.~Y.} \bibnamefont{Kupriyanov}},
  \bibnamefont{and} \bibinfo{author}{\bibfnamefont{E.}~\bibnamefont{Il'ichev}},
  \bibinfo{journal}{Rev. Mod. Phys.} \textbf{\bibinfo{volume}{76}},
  \bibinfo{pages}{411} (\bibinfo{year}{2004}).

\bibitem[{\citenamefont{de~Gennes}(1999)}]{deGennes1999}
\bibinfo{author}{\bibfnamefont{P.~G.} \bibnamefont{de~Gennes}},
  \emph{\bibinfo{title}{Superconductivity of metals and alloys}}
  (\bibinfo{publisher}{Westview Press}, \bibinfo{year}{1999}).

\bibitem[{\citenamefont{Bauch et~al.}(2005)\citenamefont{Bauch, H{\"u}rfeld,
  Krasnov, Delsing, Takayanagi, and Akazaki}}]{Bauch2005}
\bibinfo{author}{\bibfnamefont{T.}~\bibnamefont{Bauch}},
  \bibinfo{author}{\bibfnamefont{E.}~\bibnamefont{H{\"u}rfeld}},
  \bibinfo{author}{\bibfnamefont{V.~M.} \bibnamefont{Krasnov}},
  \bibinfo{author}{\bibfnamefont{P.}~\bibnamefont{Delsing}},
  \bibinfo{author}{\bibfnamefont{H.}~\bibnamefont{Takayanagi}},
  \bibnamefont{and} \bibinfo{author}{\bibfnamefont{T.}~\bibnamefont{Akazaki}},
  \bibinfo{journal}{Phys. Rev. B} \textbf{\bibinfo{volume}{71}},
  \bibinfo{pages}{174502} (\bibinfo{year}{2005}).

\bibitem[{\citenamefont{Deon et~al.}(2011)\citenamefont{Deon, Pellegrini,
  Giazotto, Biasiol, Sorba, and Beltram}}]{Deon2011}
\bibinfo{author}{\bibfnamefont{F.}~\bibnamefont{Deon}},
  \bibinfo{author}{\bibfnamefont{V.}~\bibnamefont{Pellegrini}},
  \bibinfo{author}{\bibfnamefont{F.}~\bibnamefont{Giazotto}},
  \bibinfo{author}{\bibfnamefont{G.}~\bibnamefont{Biasiol}},
  \bibinfo{author}{\bibfnamefont{L.}~\bibnamefont{Sorba}}, \bibnamefont{and}
  \bibinfo{author}{\bibfnamefont{F.}~\bibnamefont{Beltram}},
  \bibinfo{journal}{Appl. Phys. Lett.} \textbf{\bibinfo{volume}{98}},
  \bibinfo{pages}{132101} (\bibinfo{year}{2011}).

\bibitem[{\citenamefont{Dolcini and Giazotto}(2007)}]{Dolcini2007}
\bibinfo{author}{\bibfnamefont{F.}~\bibnamefont{Dolcini}} \bibnamefont{and}
  \bibinfo{author}{\bibfnamefont{F.}~\bibnamefont{Giazotto}},
  \bibinfo{journal}{Phys. Rev. B} \textbf{\bibinfo{volume}{75}},
  \bibinfo{pages}{140511} (\bibinfo{year}{2007}).

\bibitem[{\citenamefont{Alicea}(2010)}]{Alicea2010}
\bibinfo{author}{\bibfnamefont{J.}~\bibnamefont{Alicea}},
  \bibinfo{journal}{Phys. Rev. B} \textbf{\bibinfo{volume}{81}},
  \bibinfo{pages}{125318} (\bibinfo{year}{2010}).

\bibitem[{\citenamefont{Wilczek}(2009)}]{Wilczek2009}
\bibinfo{author}{\bibfnamefont{F.}~\bibnamefont{Wilczek}},
  \bibinfo{journal}{Nat. Phys.} \textbf{\bibinfo{volume}{5}},
  \bibinfo{pages}{614} (\bibinfo{year}{2009}).

\bibitem[{\citenamefont{Giazotto et~al.}(2004)\citenamefont{Giazotto,
  Grove-Rasmussen, Fazio, Beltram, Linfield, and Ritchie}}]{Giazotto2004}
\bibinfo{author}{\bibfnamefont{F.}~\bibnamefont{Giazotto}},
  \bibinfo{author}{\bibfnamefont{K.}~\bibnamefont{Grove-Rasmussen}},
  \bibinfo{author}{\bibfnamefont{R.}~\bibnamefont{Fazio}},
  \bibinfo{author}{\bibfnamefont{F.}~\bibnamefont{Beltram}},
  \bibinfo{author}{\bibfnamefont{E.~H.} \bibnamefont{Linfield}},
  \bibnamefont{and} \bibinfo{author}{\bibfnamefont{D.~A.}
  \bibnamefont{Ritchie}}, \bibinfo{journal}{J. of Supercond.}
  \textbf{\bibinfo{volume}{17}}, \bibinfo{pages}{317} (\bibinfo{year}{2004}).

\bibitem[{\citenamefont{Sch{\"a}pers et~al.}(1997)\citenamefont{Sch{\"a}pers,
  Kaluza, Neurohr, Malindretos, Crecelius, van~der Hart, Hardtdegen, and
  L{\"u}th}}]{Schapers1997}
\bibinfo{author}{\bibfnamefont{T.}~\bibnamefont{Sch{\"a}pers}},
  \bibinfo{author}{\bibfnamefont{A.}~\bibnamefont{Kaluza}},
  \bibinfo{author}{\bibfnamefont{K.}~\bibnamefont{Neurohr}},
  \bibinfo{author}{\bibfnamefont{J.}~\bibnamefont{Malindretos}},
  \bibinfo{author}{\bibfnamefont{G.}~\bibnamefont{Crecelius}},
  \bibinfo{author}{\bibfnamefont{A.}~\bibnamefont{van~der Hart}},
  \bibinfo{author}{\bibfnamefont{H.}~\bibnamefont{Hardtdegen}},
  \bibnamefont{and} \bibinfo{author}{\bibfnamefont{H.}~\bibnamefont{L{\"u}th}},
  \bibinfo{journal}{Appl. Phys. Lett.} \textbf{\bibinfo{volume}{71}},
  \bibinfo{pages}{3575} (\bibinfo{year}{1997}).

\bibitem[{\citenamefont{Franceschi et~al.}(2010)\citenamefont{Franceschi,
  Kouwenhoven, Sch{\"o}nenberger, and Wernsdorfer}}]{DeFranceschi2010}
\bibinfo{author}{\bibfnamefont{S.~D.} \bibnamefont{Franceschi}},
  \bibinfo{author}{\bibfnamefont{L.}~\bibnamefont{Kouwenhoven}},
  \bibinfo{author}{\bibfnamefont{C.}~\bibnamefont{Sch{\"o}nenberger}},
  \bibnamefont{and}
  \bibinfo{author}{\bibfnamefont{W.}~\bibnamefont{Wernsdorfer}},
  \bibinfo{journal}{Nature Nanotech.} \textbf{\bibinfo{volume}{5}},
  \bibinfo{pages}{703} (\bibinfo{year}{2010})
  \bibinfo{note}{$\!\!$, and references therein}.

\bibitem[{\citenamefont{Desrat et~al.}(2004)\citenamefont{Desrat, Giazotto,
  Pellegrini, Beltram, Capotondi, Biasiol, Sorba, and Maude}}]{Desrat2004}
\bibinfo{author}{\bibfnamefont{W.}~\bibnamefont{Desrat}},
  \bibinfo{author}{\bibfnamefont{F.}~\bibnamefont{Giazotto}},
  \bibinfo{author}{\bibfnamefont{V.}~\bibnamefont{Pellegrini}},
  \bibinfo{author}{\bibfnamefont{F.}~\bibnamefont{Beltram}},
  \bibinfo{author}{\bibfnamefont{F.}~\bibnamefont{Capotondi}},
  \bibinfo{author}{\bibfnamefont{G.}~\bibnamefont{Biasiol}},
  \bibinfo{author}{\bibfnamefont{L.}~\bibnamefont{Sorba}}, \bibnamefont{and}
  \bibinfo{author}{\bibfnamefont{D.~K.} \bibnamefont{Maude}},
  \bibinfo{journal}{Phys. Rev. B} \textbf{\bibinfo{volume}{69}},
  \bibinfo{pages}{245324} (\bibinfo{year}{2004}).

\bibitem[{\citenamefont{Flensberg et~al.}(1988)\citenamefont{Flensberg, Hansen,
  and Octavio}}]{Flensberg1988}
\bibinfo{author}{\bibfnamefont{K.}~\bibnamefont{Flensberg}},
  \bibinfo{author}{\bibfnamefont{J.~B.} \bibnamefont{Hansen}},
  \bibnamefont{and} \bibinfo{author}{\bibfnamefont{M.}~\bibnamefont{Octavio}},
  \bibinfo{journal}{Phys. Rev. B} \textbf{\bibinfo{volume}{38}},
  \bibinfo{pages}{8708} (\bibinfo{year}{1988}).

\bibitem[{\citenamefont{Courtois et~al.}(2008)\citenamefont{Courtois, Meschke,
  Peltonen, and Pekola}}]{Courtois2008}
\bibinfo{author}{\bibfnamefont{H.}~\bibnamefont{Courtois}},
  \bibinfo{author}{\bibfnamefont{M.}~\bibnamefont{Meschke}},
  \bibinfo{author}{\bibfnamefont{J.}~\bibnamefont{Peltonen}}, \bibnamefont{and}
  \bibinfo{author}{\bibfnamefont{J.}~\bibnamefont{Pekola}},
  \bibinfo{journal}{Phys. Rev. Lett} \textbf{\bibinfo{volume}{101}},
  \bibinfo{pages}{057005} (\bibinfo{year}{2008}).

\bibitem[{\citenamefont{Dubos et~al.}(2001)\citenamefont{Dubos, Courtois,
  Pannetier, Wilhelm, Zaikin, and Sch{\"o}n}}]{Dubos2001}
\bibinfo{author}{\bibfnamefont{P.}~\bibnamefont{Dubos}},
  \bibinfo{author}{\bibfnamefont{H.}~\bibnamefont{Courtois}},
  \bibinfo{author}{\bibfnamefont{B.}~\bibnamefont{Pannetier}},
  \bibinfo{author}{\bibfnamefont{F.~K.} \bibnamefont{Wilhelm}},
  \bibinfo{author}{\bibfnamefont{A.~D.} \bibnamefont{Zaikin}},
  \bibnamefont{and}
  \bibinfo{author}{\bibfnamefont{G.}~\bibnamefont{Sch{\"o}n}},
  \bibinfo{journal}{Phys. Rev. B} \textbf{\bibinfo{volume}{63}},
  \bibinfo{pages}{064502} (\bibinfo{year}{2001}).

\bibitem{note1}
  \bibinfo{note}{We estimate the uncertainty in the evaluation of $I_s$ to be of the order of $\sim 10\%$ around \SI{1.1}{K}, $\sim 25\%$ around \SI{1.5}{K} and $\sim 100\%$ around \SI{2.0}{K}.}

\bibitem[{\citenamefont{Hammer et~al.}(2007)\citenamefont{Hammer, Cuevas,
  Bergeret, and Belzig}}]{Hammer2007}
\bibinfo{author}{\bibfnamefont{J.~C.} \bibnamefont{Hammer}},
  \bibinfo{author}{\bibfnamefont{J.~C.} \bibnamefont{Cuevas}},
  \bibinfo{author}{\bibfnamefont{F.~S.} \bibnamefont{Bergeret}},
  \bibnamefont{and} \bibinfo{author}{\bibfnamefont{W.}~\bibnamefont{Belzig}},
  \bibinfo{journal}{Phys. Rev. B} \textbf{\bibinfo{volume}{76}},
  \bibinfo{pages}{064514} (\bibinfo{year}{2007}).

\bibitem[{\citenamefont{Heikkil{\"a} et~al.}(2002)\citenamefont{Heikkil{\"a},
  S{\"a}rkk{\"a}, and Wilhelm}}]{Heikkila2002}
\bibinfo{author}{\bibfnamefont{T.~T.} \bibnamefont{Heikkil{\"a}}},
  \bibinfo{author}{\bibfnamefont{J.}~\bibnamefont{S{\"a}rkk{\"a}}},
  \bibnamefont{and} \bibinfo{author}{\bibfnamefont{F.~K.}
  \bibnamefont{Wilhelm}}, \bibinfo{journal}{Phys. Rev. B}
  \textbf{\bibinfo{volume}{66}}, \bibinfo{pages}{184513}
  (\bibinfo{year}{2002}).

\bibitem[{\citenamefont{Tinkham}(1996)}]{Tinkham1996}
\bibinfo{author}{\bibfnamefont{M.}~\bibnamefont{Tinkham}},
  \emph{\bibinfo{title}{Introduction to Superconductivity}}
  (\bibinfo{year}{1996}), \bibinfo{note}{2nd edition}.

\bibitem[{\citenamefont{Gu et~al.}(1979)\citenamefont{Gu, Cha, and
  Gamo}}]{Gu1979}
\bibinfo{author}{\bibfnamefont{J.}~\bibnamefont{Gu}},
  \bibinfo{author}{\bibfnamefont{W.}~\bibnamefont{Cha}}, \bibnamefont{and}
  \bibinfo{author}{\bibfnamefont{K.}~\bibnamefont{Gamo}}, \bibinfo{journal}{J.
  Appl. Phys.} \textbf{\bibinfo{volume}{50}}, \bibinfo{pages}{6437}
  (\bibinfo{year}{1979}).

\bibitem[{\citenamefont{Heida et~al.}(1998)\citenamefont{Heida, van Wees,
  Klapwijk, and Borghs}}]{Heida1998}
\bibinfo{author}{\bibfnamefont{J.~P.} \bibnamefont{Heida}},
  \bibinfo{author}{\bibfnamefont{B.~J.} \bibnamefont{van Wees}},
  \bibinfo{author}{\bibfnamefont{T.~M.} \bibnamefont{Klapwijk}},
  \bibnamefont{and} \bibinfo{author}{\bibfnamefont{G.}~\bibnamefont{Borghs}},
  \bibinfo{journal}{Phys. Rev. B} \textbf{\bibinfo{volume}{57}},
  \bibinfo{pages}{R5618} (\bibinfo{year}{1998}).

\bibitem[{\citenamefont{Barzykin and Zagoskin}(1999)}]{Barzykin1999}
\bibinfo{author}{\bibfnamefont{V.}~\bibnamefont{Barzykin}} \bibnamefont{and}
  \bibinfo{author}{\bibfnamefont{A.}~\bibnamefont{Zagoskin}},
  \bibinfo{journal}{Superlatt. and Microstruct.} \textbf{\bibinfo{volume}{25}},
  \bibinfo{pages}{797} (\bibinfo{year}{1999}).

\bibitem[{\citenamefont{Grajcar et~al.}(2002)\citenamefont{Grajcar, Ebel,
  Il'ichev, K{\"u}rsten, Matsuyama, and Merkt}}]{Grajcar2002}
\bibinfo{author}{\bibfnamefont{M.}~\bibnamefont{Grajcar}},
  \bibinfo{author}{\bibfnamefont{M.}~\bibnamefont{Ebel}},
  \bibinfo{author}{\bibfnamefont{E.}~\bibnamefont{Il'ichev}},
  \bibinfo{author}{\bibfnamefont{R.}~\bibnamefont{K{\"u}rsten}},
  \bibinfo{author}{\bibfnamefont{T.}~\bibnamefont{Matsuyama}},
  \bibnamefont{and} \bibinfo{author}{\bibfnamefont{U.}~\bibnamefont{Merkt}},
  \bibinfo{journal}{Physica C} \textbf{\bibinfo{volume}{372}},
  \bibinfo{pages}{27} (\bibinfo{year}{2002}).

\bibitem{note2}
  \bibinfo{note}{We are assuming that the charge density in the 2DEG surrounding the QD can be still assumed uniform and equal to the value in the open configuration.}

\bibitem[{\citenamefont{Levy-Yeyati et~al.}(1997)\citenamefont{Levy-Yeyati,
  Cuevas, Lopez-Davalos, and Mart{\'i}n-Rodero}}]{Yeyati1997}
\bibinfo{author}{\bibfnamefont{A.}~\bibnamefont{Levy-Yeyati}},
  \bibinfo{author}{\bibfnamefont{J.~C.} \bibnamefont{Cuevas}},
  \bibinfo{author}{\bibfnamefont{A.}~\bibnamefont{Lopez-Davalos}},
  \bibnamefont{and}
  \bibinfo{author}{\bibfnamefont{A.}~\bibnamefont{Mart{\'i}n-Rodero}},
  \bibinfo{journal}{Phys. Rev. B} \textbf{\bibinfo{volume}{55}},
  \bibinfo{pages}{R6137} (\bibinfo{year}{1997}).

\bibitem[{\citenamefont{Dynes et~al.}(1978)\citenamefont{Dynes, Narayanamurti,
  and Garno}}]{Dynes1978}
\bibinfo{author}{\bibfnamefont{R.~C.} \bibnamefont{Dynes}},
  \bibinfo{author}{\bibfnamefont{V.}~\bibnamefont{Narayanamurti}},
  \bibnamefont{and} \bibinfo{author}{\bibfnamefont{J.~P.} \bibnamefont{Garno}},
  \bibinfo{journal}{Phys. Rev. Lett} \textbf{\bibinfo{volume}{41}},
  \bibinfo{pages}{1509} (\bibinfo{year}{1978}).

\end{thebibliography}

\end{document}